\documentclass[english,aps,prl,superscriptaddress,twocolumn]{revtex4-1}
 \pdfoutput=1
\usepackage[T1]{fontenc}
\usepackage[latin9]{inputenc}
\usepackage{amsmath}
\usepackage{graphicx}
\usepackage{esint}
\usepackage{xcolor}

\makeatletter
\usepackage{slashed}
\usepackage{hyperref}

\makeatother

\usepackage{babel}
\begin{document}

\title{Nieh-Yan Anomaly: Torsional Landau Levels, central charge and anomalous thermal Hall effect}

\author{Ze-Min Huang}
\email{zeminh2@illinois.edu}

\affiliation{Department of Physics, University of Illinois at Urbana-Champaign,
1110 West Green Street, Urbana, Illinois 61801, USA}

\author{Bo Han}

\affiliation{Theory of Condensed Matter Group, Cavendish Laboratory, University of Cambridge,
J.~J.~Thomson Avenue, Cambridge CB3 0HE,
United Kingdom}

\affiliation{Department of Physics, University of Illinois at Urbana-Champaign,
1110 West Green Street, Urbana, Illinois 61801, USA}

\author{Michael Stone}

\affiliation{Department of Physics, University of Illinois at Urbana-Champaign,
1110 West Green Street, Urbana, Illinois 61801, USA}

\date{\today}

\begin{abstract}
The Nieh-Yan anomaly is the anomalous breakdown of the chiral U(1) symmetry
caused by the interaction between torsion and fermions. We study this
anomaly from the point of view of torsional Landau levels. It was
found that the torsional Landau levels are gapless, while their contributions
to the chiral anomaly are canceled, except those from the lowest torsional
Landau levels. Hence, the dimension is effectively reduced from (3+1)-dimensional to (1+1)-dimensional.
We further show that the coefficient of the Nieh-Yan anomaly is the
free energy density in (1+1) dimensions. Especially, at finite temperature, the thermal
Nieh-Yan anomaly is proportional to the central charge. The anomalous
thermal Hall conductance in Weyl semimetals is then shown to be proportional
to the central charge, which is the experimental fingerprint of the
thermal Nieh-Yan  anomaly.
\end{abstract}

\maketitle

\section{Introduction}

Although torsion naturally arises as the curvature tensor of the translational
gauge fields \citep{hehl1976rmp}, in gravity, its observable
effects are relatively small and are usually neglected. By contrast,
torsion is attracting more and more attention in condensed matter
physics. Torsion can emerge from dislocations \citep{katanaev1992aop,taylor2011prl,taylor2013prd, onkar2014prd},
the temperature gradient \citep{luttinger1964pr,shitade2014ptep,tatara2015prl,barry2015prb}, background rotation and the order parameter of Fermi superfluid or topological superconductors \citep{volovik2003oxford,golan2018prb,Nissinen2019arxivNY}. Especially, in Dirac and Weyl semimetals,
due to their gapless spectrum and strong spin-orbit coupling \citep{armitage2018rmp,burkov2018arcp},
torsion has led to rich physical phenomena, for example, chiral zero
modes trapped in dislocations \citep{chernodub2017prb,huang2019prb},
the chiral torsional magnetic effect \citep{sumiyoshi2016prl} and
other viscoelastic responses \citep{sun2014epl,you2016prb,palumbo2016aop,khaidukov2018jetp,ferreiros2019prl}.

Topological phases of matter are closely related to quantum anomalies
\citep{ryu2012prb,witten2016rmp}. Similarly, both the chiral anomaly
and mixed axial gravitational anomaly are important to understand Dirac and Weyl semimetals \citep{vilenkin1979prd,vilenkin1980prd,fukushima2008prd,landsteiner2011prl,zyuzin2012prb,stephanov2012prl,zhou2013cpl,li2015naturecom,spivak2016prb,li2016naturecom,li2016naturephy,jia2016nc,gorbar2017prbBZ,huang2017prb,stone2018prd}.
Torsion can lead to chiral current non-conservation as well, which
is known as the Nieh-Yan anomaly \citep{nieh1982aop,nieh1982jmp}.
However, compared to other anomalies, the Nieh-Yan anomaly depends
on the cutoff and thus the specific ultra-violet physics, which is
still controversial \citep{chandia2001prd,kreimer2001prd}. The
Nieh-Yan anomaly lies in the intersection of condensed matter physics
and high-energy physics, topology and geometry. Meanwhile, Dirac
and Weyl semimetals have provided an ideal platform to study these phenomena
on a tabletop system.

Recently, it was suggested in Ref. \citep{nissinen2019arxiv} that
there might be an extra thermal term in the Nieh-Yan anomaly, i.e.,
\begin{equation}
\frac{1}{\sqrt{|g|}}\partial_{\mu}\sqrt{|g|}j^{5\mu}=\left(\frac{\Lambda^{2}}{4\pi^{2}}-\frac{T^{2}}{12}\right)\frac{\epsilon^{\mu\nu\rho\sigma}}{\sqrt{|g|}}\partial_{\mu}e_{\nu}^{a}\partial_{\rho}e_{\sigma}^{a}\eta_{ab},\label{eq:thermal_Nieh_Yan}
\end{equation}
where $g_{\mu\nu}$ is the metric and $g=\det{g_{\mu\nu}}$. $\Lambda$ is the cutoff, $T$ is the temperature, $e_{\mu}^{a}$
is the vierbeins and for simplicity, the spin connection is set to
zero. From the view of Poincare gauge theory, this anomaly has
a form similar to the Adler-Bell-Jackiw anomaly \citep{adler1969pr,bell1969nca}.
But what is the physical meaning and physical mechanism behind this
anomaly? Can we understand this anomaly equation from any kind of
``Landau levels'' \citep{nielsen1983plb}? Compared to the zero-temperature
Nieh-Yan term, the coefficient of the $T^{2}$ term is dimensionless,
so it is also tempting to ask if they are related to any kind of
topological invariant and why the coefficient is $\frac{T^{2}}{12}$.

In this paper, we derive the energy spectrum for Weyl fermions under
torsional magnetic fields. The energy spectrum turns out to be sharply different from its magnetic counterparts. Namely, all the Landau
levels collapse together at the zero-momentum point. Interestingly,
only the lowest torsional Landau levels matter as far as the axial
currents are concerned, because the effect from the higher torsional
Landau levels is canceled exactly. Thus, the $\left(3+1\right)$-dimensional
system is effectively reduced to pairs of $\left(1+1\right)$-dimensional Dirac
fermions. Especially, the axial current is proportional to the free
energy density of the effective $\left(1+1\right)$-dimensional Dirac fermions.
If the torsional electric fields are further applied, the effective
velocity of the lowest torsional Landau levels is changed. Since
the free energy density of $\left(1+1\right)$-dimensional conformal fields
is proportional to central charge as well as the inverse of velocity
\citep{affleck1986prl,blote1986prl}, the torsional electric fields
change the axial current and thus lead to chiral anomaly. Hence,
we have found that the coefficient of the thermal Nieh-Yan anomaly
in Eq. (\ref{eq:thermal_Nieh_Yan}) is
\begin{equation}
\frac{T^{2}}{12}=\left(\frac{\pi c}{6}T^{2}\right)\left(\frac{1}{2\pi}\right),
\end{equation}
where $\frac{\pi c}{6}T^{2}$ is the free energy density of some $\left(1+1\right)$-dimensional
conformal fields (the lowest torsional Landau levels), $\frac{1}{2\pi}$
is from the level degeneracy, and $c=1$ is the central charge. The
$\Lambda^{2}$ term in Eq. (\ref{eq:thermal_Nieh_Yan}) is from the vacuum energy, for example, the band
depth in Weyl semimetals, so it depends on concrete materials. By
contrast, the thermal term is proportional to the central charge,
which is universal. Finally, we show that the anomalous thermal Hall 
effect in Weyl semimetals is induced by the thermal Nieh-Yan anomaly,
which serves as the experimental signature of the thermal Nieh-Yan
anomaly.

The rest of this paper is organized as follow. In Sec.~ \ref{sec:Landau_level},
the energy spectrum for Weyl fermions under torsional magnetic fields
is derived. In Sec.~\ref{sec:Nieh_Yan_levels}, both the chiral torsional
effect and the Nieh-Yan anomaly are calculated from the lowest torsional
Landau levels. In Sec.~\ref{sec:effective_model}, an effective model is
constructed from the lowest torsional Landau levels, which relates
the thermal Nieh-Yan anomaly  to the central charge. In Sec.~ \ref{sec:responses},
we derive the anomalous thermal Hall effect in Weyl semimetals from
the thermal Nieh-Yan anomaly. We summarize the main results of this paper in Sec.~\ref{sec:conclusion}.

\begin{figure}
$a)$\includegraphics[scale=0.5]{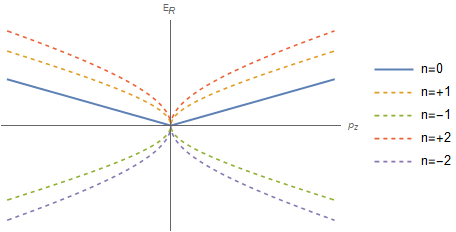}

$b)$\includegraphics[scale=0.5]{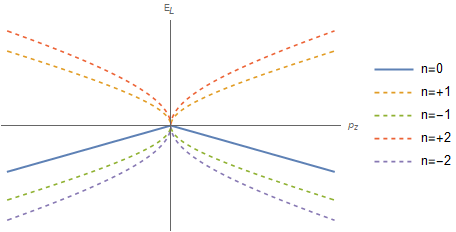}

\caption{Torsional Landau levels. Top: the energy spectrum for right-handed
Weyl fermions under torsional magnetic fields. Bottom: the energy
spectrum for left-handed Weyl fermions. The energy for the right-handed
lowest torsional Landau level is positive, while it is negative for the
left-handed one. \label{fig:Torsional_LL} }
\end{figure}

\section{torsional magnetic fields and Torsional Landau levels \label{sec:Landau_level}}

Let us consider the following action for Weyl fermions in curved spacetime,

\begin{equation}
S= \int d^{d}xe\frac{1}{2}\left[\bar{\psi}e_{a}^{\mu}\gamma^{a}\left(i\partial_{\mu}\right)\psi-\bar{\psi}\left(i\overleftarrow{\partial}_{\mu}\right)e_{a}^{\mu}\gamma^{a}\psi\right],
\end{equation}
where the spin connection is set to zero and $a,\thinspace\mu=0,\thinspace1,\thinspace2,\thinspace3$
denote the locally flat coordinates (coordinate vectors $e_{a}^{\mu}\partial_{\mu}$)
and curved coordinates, respectively. $\gamma^{a}$ is the gamma matrix, i.e., $\gamma^0=\sigma^{0}\otimes \tau^{1}$ and $\gamma^i=\sigma^{i}\otimes (-i\tau^2)$, where both $\sigma$ and $\tau$ stand for the Pauli matrices. The vierbein $e_{a}^{\mu}$
and its inverse $e_{\nu}^{a}$ satisfy $e_{a}^{\mu}e_{\mu}^{b}=\delta_{a}^{b}$.
In addition, these vierbeins can be realized in Weyl semimetals, for
example, by dislocation \citep{katanaev1992aop,taylor2011prl,taylor2013prd,onkar2014prd},
temperature gradient \citep{luttinger1964pr,tatara2015prl} and global
rotation.

Now for simplicity, let us consider a specific configuration of the
vierbeins, namely, $e_{\mu}^{a}=\delta_{\mu}^{a}+w_{\mu}^{a}$ and
$w_{\mu}^{a}=\frac{1}{2}\delta_{3}^{a}\tilde{T}_{B}^{3}\left(0,\thinspace-y,\thinspace x,\thinspace0\right),\text{\ensuremath{\tilde{T}_{B}^{3}>0}}$,
which means that the torsional magnetic fields are applied along the
$z$ direction. The corresponding Hamiltonian is

\begin{eqnarray}
H_{s} & = & s[p_{z}\sigma^{3}+\left(\hat{p}_{x}+\frac{1}{2}\tilde{T}_{B}^{3}y p_{z}\right)\sigma^{1}\nonumber \\
 &  & +\left(\hat{p}_{y}-\frac{1}{2}\tilde{T}_{B}^{3}x p_{z}\right)\sigma^{2}]
\end{eqnarray}
where $p_{z}$ is a good quantum number and $s=\pm1$ denotes the
chirality. Compared to the magnetic case, this Hamiltonian looks like
Weyl fermions under magnetic fields with charge $p_{z}$. The dispersion
relation of this Hamiltonian can be straightforwardly derived, i.e.,

\begin{equation}
\mathcal{E}_{s}=\begin{cases}
\begin{array}{c}
s\left|p_{z}\right|\\
\pm\sqrt{p_{z}^{2}+2\left|n\tilde{T}_{B}^{3}p_{z}\right|}
\end{array} & \begin{array}{c}
n=0\\
|n|\geq1
\end{array}\end{cases},
\end{equation}
where the level degeneracy is $\frac{1}{2\pi}\left|p_{z}\right|\tilde{T}_{B}^{3}$
and the spectrum is shown in Fig. \ref{fig:Torsional_LL}. The energy
spectrum with $\left|n\right|\geq1$ is the same for Weyl fermions
with different chiralities. So only the lowest torsional Landau levels
can distinguish fermions with different chiralities. However, compared
to the magnetic Landau levels, there are two main differences. First,
all of the Landau levels collapse together at $p_{z}=0$ , which is because
the torsional magnetic charge $p_{z}$ vanishes at this point. Second,
the lowest torsional Landau level is of the form $s\left|p_{z}\right|$
rather than $sp_{z}$ in the magnetic case, which is because $p_{z}$
reverse its sign at $p_{z}=0$.

\begin{figure}
\includegraphics[scale=0.45]{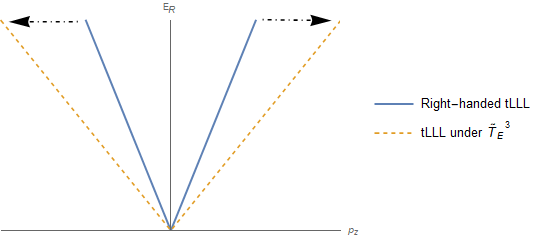}\caption{An illustration of the Nieh-Yan anomaly under torsional electric fields
$\tilde{T}_{E}=\partial_{t}\Phi$ (``tLLL'' stands for the torsional
lowest Landau levels). The torsional electric fields change the slope
of the lowest Landau level, for example, from the thick line (velocity
$v=\text{\ensuremath{1}}$) to the dashed line [velocity $v=\text{\ensuremath{\left(1-\Phi\right)}}$].
Because the free energy density of two-dimensional conformal
field theory is $\frac{\pi cT^{2}}{6}\frac{1}{v}$ \citep{affleck1986prl,blote1986prl},
the change in chiral current density is $\Delta j^{50}=\left(\frac{1}{v^{2}}-1\right)\left(\frac{\pi cT^{2}}{6}\right)\left(\frac{\tilde{T}_{B}^{3}}{2\pi}\right)$, where  $\frac{\tilde{T}_{B}^{3}}{2\pi}$ is from the level degeneracy. Thus, $\frac{\Delta j^{50}}{\Delta t}=\frac{c}{6}T^{2}\tilde{T}_{E}^{3}\tilde{T}_{B}^{3}$.
 \label{fig:illustration_Nieh-Yan}}
\end{figure}

\section{Nieh-Yan term from the Lowest Torsional Landau levels \label{sec:Nieh_Yan_levels}}

After deriving the torsional Landau levels, it is natural to ask if
we can extract the Nieh-Yan anomaly from these levels and what
the physical meaning of this anomaly is. In this section, we shall show
that regardless of the gapless nature in the energy spectrum, the
higher torsional Landau levels ($\left|n\right|\geq1$) do not contribute
to the anomaly equation. Thus, the Nieh-Yan anomaly arises from the
lowest torsional Landau levels and the system is effectively reduced
to $\left(1+1\right)$-dimensional Dirac fermions. In addition, the
prefactor of the Nieh-Yan term is the free energy density of the effectively
$\left(1+1\right)$-dimensional Dirac fermions.

Now we further turn on the torsional electric fields, i.e., $\tilde{T}_{E}^{3}=\partial_{0}e_{z}^{3}-\partial_{z}e_{0}^{3}$.
For simplicity, we set $e_{0}^{3}=0$ and $e_{z}^{3}=1+\Phi$, $\Phi\ll1$.
Due to the ``minimal coupling'', i.e., $\sigma^{a}\rightarrow e_{a}^{\mu}\sigma^{a}$,
only $\sigma^{3}$ in the Hamiltonian are modified and the dispersion
relation now becomes
\begin{equation}
\mathcal{E}_{s}=\begin{cases}
\begin{array}{cc}
s\left(1-\Phi\right)\left|p_{z}\right| & n=0\\
\pm\sqrt{\left|\left(1-\Phi\right)p_{z}\right|^{2}+2\left|n\tilde{T}_{B}^{3}p_{z}\right|} & n^{2}\geq1
\end{array} & ,\end{cases}
\end{equation}
where for $e_{\mu}^{a}=\delta_{\mu}^{a}+w_{a}^{\mu}$, $e_{a}^{\mu}\simeq\delta_{a}^{\mu}-\delta_{a}^{\nu}\delta_{b}^{\mu}w_{\nu}^{b}$
and  $e_{3}^{z}\simeq1-\Phi$. This shows that the torsional electric
fields affect the lowest Landau levels by modifying their slope (or
velocity), i.e., $\left|p_{z}\right|\rightarrow\left(1-\Phi\right)\left|p_{z}\right|$.

The axial charge density $j^{5\mu}|_{\mu=0}$ can be written as
\begin{equation}
j^{50}=\sum_{n}\sum_{s}s\int_{-\infty}^{+\infty}\frac{dp_{z}}{2\pi}n_{F}\left(\mathcal{E}_{s}^{n}\right)\left(\frac{\left|p_{z}\right|\tilde{T}_{B}^{3}}{2\pi}\right),
\end{equation}
where $n=0,\thinspace\pm1,\thinspace\dots$ is used to label the torsional
Landau levels, $n_{F}\left(\mathcal{E}^{n}_{s}\right)=\frac{1}{\exp\left(\beta\mathcal{E}^{n}_{s}\right)+1}$
is the Fermi-Dirac distribution function and $\frac{\left|p_{z}\right|\tilde{T}_{B}^{3}}{2\pi}$
is the level degeneracy. Because that for $\left|n\right|\geq1$,
$\mathcal{E}_{R}^{n}=\mathcal{E}_{L}^{n}$, there is $\sum_{\left|n\right|>1}\sum_{s}sn_{F}\left(\mathcal{E}_{s}^{n}\right)\left|p_{z}\right|=0$,
so the contributions from the higher torsional Landau levels ($|n|\geq1$) to $j^{50}$ are canceled exactly.

Now the axial charge density becomes

\begin{eqnarray}
 &  & \left(\sum_{s}sj_{s}^{0}\right)/\left[\tilde{T}_{B}^{3}/\left(2\pi\right)\right]\nonumber \\
 & = & \int_{-\infty}^{+\infty}\frac{dp_{z}}{2\pi}\left|p_{z}\right|\nonumber \\
 &  & \times\left\{ \frac{1}{\exp\left[\beta\left(1-\Phi\right)\left|p_{z}\right|\right]+1}-\frac{1}{\exp\left[-\beta\left(1-\Phi\right)\left|p_{z}\right|\right]+1}\right\} \nonumber \\
 & = & \frac{1}{\left(1-\Phi\right)^{2}}2\int_{-\infty}^{+\infty}\frac{d\epsilon}{2\pi}\epsilon\frac{1}{\exp\left(\beta\epsilon\right)+1},\label{eq:1_dim_anomaly}
\end{eqnarray}
where $\epsilon\equiv\left(1-\Phi\right)\left|p_{z}\right|$. If we
regard $\left(1-\Phi\right)$ as the effective velocity, then terms
in the second line are almost the energy density of the $\left(1+1\right)$-dimensional
Dirac fermions. By ``almost,'' we mean that the coefficient of $\left|p_{z}\right|$
in the integrand is not $\left(1-\Phi\right)$. In the last line,
$2\int_{-\infty}^{+\infty}\frac{d\epsilon}{2\pi}\epsilon\frac{1}{\exp\left(\beta\epsilon\right)+1}$
is the energy density of two different kinds of fermions. This suggests
the close relation between the Nieh-Yan anomaly and the energy density.
It is also known that in two dimensional conformal field theory, the
free energy density is proportional to the central charge \citep{affleck1986prl,blote1986prl},
so this equation also hints at the close relation between the Nieh-Yan
anomaly and the central charge. This connection will be explored
in the next section.

However, Eq. (\ref{eq:1_dim_anomaly}) is actually divergent, so we
need to perform the integration carefully. To be more concrete, we
rewrite Eq. (\ref{eq:1_dim_anomaly}) as

\begin{equation}
\frac{2}{\left(1-\Phi\right)^{2}}
\left\{ \int_{-\infty}^{+\infty}\frac{d\epsilon}{2\pi}\epsilon\left[\frac{1}{\exp\left(\beta\epsilon\right)+1}-\theta\left(-\epsilon\right)\right]+\int_{-\infty}^{+\infty}\frac{d\epsilon}{2\pi}\epsilon\theta\left(-\epsilon\right)\right\} .
\end{equation}
The first term in the bracket is now convergent, i.e.,
\[
\int_{-\infty}^{+\infty}\frac{d\epsilon}{2\pi}\frac{\epsilon}{2\pi}\left[\frac{1}{\exp\left(\beta\epsilon\right)+1}-\theta\left(-\epsilon\right)\right]=\frac{T^{2}}{4!}.
\]
But the second term can be regularized by introducing a hard energy cut-off, i.e., $-\Lambda<\epsilon<\Lambda$. Then, the integral becomes
\[
\int_{-\Lambda}^{\Lambda}\frac{d\epsilon}{2\pi}\epsilon\theta\left(-\epsilon\right)=-\frac{1}{4\pi}\Lambda^{2},
\]
where the positive $\epsilon$ part in the integral is eliminated
by $\theta\left(-\epsilon\right)$ and the negative part is regularized
by $-\Lambda$. Hence, $\Lambda$ measures the depth of the vacuum
and it stands for the vacuum energy. In reality, the depth of the
vacuum is not universal. For example, it might depend on the concrete
materials. By contrast, the term proportional to $T^{2}$ can be universal.
As we shall show in the next section, it is proportional to the central
charge.

By summing everything together, 
\begin{equation}
j^{50}=\left(\frac{1}{12}T^{2}-\frac{1}{4\pi^{2}}\Lambda^{2}\right)\tilde{T}_{B}^{3}+\left(\frac{1}{6}T^{2}-\frac{1}{2\pi^{2}}\Lambda^{2}\right)\Phi\tilde{T}_{B}^{3}.\label{eq:chiral_current}
\end{equation}
 The first term in the parentheses is the chiral torsional effect obtained
in Ref. \citep{khaidukov2018jetp}. In addition to the thermal energy,
the vacuum energy can affect the chiral torsional effect as well,
which leads to the $\Lambda^{2}$ term.

We can also study the chiral anomaly from Eq. (\ref{eq:chiral_current}).
For example, by recasting $\partial_{t}j^{50}$ in a covariant form,

\begin{equation}
\frac{1}{\sqrt{|g|} }\partial_{\mu}\sqrt{|g|}j^{5\mu}=  \left(-\frac{T^{2}}{12}+\frac{\Lambda^{2}}{4\pi^{2}}\right)\frac{\epsilon^{\mu\nu\rho\sigma}}{\sqrt{|g|}}\partial_{\mu}e_{\nu}^{a}\partial_{\rho}e_{\sigma}^{b}\eta_{ab},
\end{equation}
which is the Nieh-Yan term with a thermal contribution. The term proportional
to $T^{2}$ can be understood as follow. Under torsional electric
fields, the velocity of the lowest torsional Landau levels is changed.
For example, in Fig. \ref{fig:illustration_Nieh-Yan}, the velocities
(or slopes) of the thick solid line and dashed line are $v=1$ and $v=\left(1-\Phi\right)$,
respectively, where $\Phi=\tilde{T}_{E}^{3}\Delta t$. It is known
that for two dimensional conformal fields, the free energy density
at velocity $v$ is $\mathcal{F}\left(v\right)=\frac{\pi c}{6}T^{2}\frac{1}{v}$
\citep{affleck1986prl,blote1986prl}, where $c$ is the central charge
and $c=1$ for the $\left(1+1\right)$-dimensional Dirac fermions here. In
Eq. (\ref{eq:1_dim_anomaly}), the chiral density is $\left[\frac{1}{v}\times\mathcal{F}\left(v\right)\right]\frac{\tilde{T}_{B}^{3}}{2\pi}$,
where the $\frac{1}{v}$ is because the torsional Landau level
degeneracy is not affected by the torsional electric fields. Thus,
the change in chiral density is $\Delta j^{50}=\left[\frac{1}{\left(1-\Phi\right)^{2}}-1\right]\left(\frac{\pi c}{6}T^{2}\right)\left(\frac{\tilde{T}_{B}^{3}}{2\pi}\right)$
or $\Delta j^{50}\simeq\frac{c}{6}T^{2}\tilde{T}_{E}^{3}\tilde{T}_{B}^{3}\Delta t$,
with $c=1$.

This implies that the prefactor of the Nieh-Yan term is actually the
free energy density. Especially, the coefficient of the first term
is from
\begin{equation}
\frac{T^{2}}{12}=\left(\frac{c\pi T^{2}}{6}\right)\left(\frac{1}{2\pi}\right),
\end{equation}
 where $c=1$, $\frac{1}{2\pi}$ is from the level degeneracy and
$\frac{c\pi T^{2}}{6}$ is the free energy density. This strongly
suggests the close relation between the central charge and the Nieh-Yan
anomaly.

\section{The $\left(1+1\right)$-Dimensional Effective model and Central charge \label{sec:effective_model}}

In the last section, we derived both the Nieh-Yan anomaly and the
chiral torsional effect from the lowest torsional Landau levels, which
seem to relate to the central charge. In this section, we shall study
this connection by projecting the $\left(3+1\right)$-dimensional system onto
its lowest torsional Landau levels, which is effectively the $\left(1+1\right)$-dimensional
Dirac theory.

Since only the lowest Landau levels contribute to the anomaly equation,
we can project our $\left(3+1\right)$-dimensional system onto the lowest
torsional Landau levels. Then, the effective Lagrangian is

\begin{equation}
\mathcal{L}=\psi^{\dagger}\left(i\partial_{t}+\left|p_{z}\right|\sigma^{3}\right)\psi.
\end{equation}
The corresponding chiral current defined in two dimensions is
\begin{equation}
j_{c}^{\mu}=\bar{\psi}\Gamma^{\mu}\Gamma^{5}\psi,\label{eq:chiral_1+1}
\end{equation}
where $\Gamma^{\mu}$ and $\Gamma^{5}$ are the gamma matrices defined
on the $\left(1+1\right)$-dimensional spacetime.  We can further recast the
Lagrangian above as
\begin{equation}
\mathcal{L}=\bar{\psi^{\prime}}\left(i\partial_{\mu}\Gamma^{\mu}\right)\psi^{\prime},
\end{equation}
where $\psi^{\prime}\left(p_{z}\right)=\left(\begin{array}{c}
\psi_{R}\\
\psi_{L}
\end{array}\right)$ for $p_{z}>0$ and $\psi^{\prime}\left(p_{z}\right)=\left(\begin{array}{c}
\psi_{L}\\
\psi_{R}
\end{array}\right)$ for $p_{z}<0$.

Hence, in terms of $\psi^{\prime}$, Eq. (\ref{eq:chiral_1+1}) can
be written as $j_{c}^{\mu}\left(p_{z}\right)=\text{\text{sgn}\ensuremath{\left(p_{z}\right)}\ensuremath{\ensuremath{\bar{\psi^{\prime}}}}\ensuremath{\Gamma^{\mu}}\ensuremath{\Gamma^{5}}\ensuremath{\psi^{\prime}}}.$
Notice that the level degeneracy of the torsional Landau level is
$\frac{1}{2\pi}\tilde{T}_{B}^{3}\left|p_{z}\right|$, so the actual
chiral current in the $\left(3+1\right)$-dimensional spacetime is
$j^{5\mu}=\left(\frac{\tilde{T}_{B}^{3}}{2\pi}\right)\left(\bar{\psi^{\prime}}\Gamma^{\mu}\Gamma^{5}p_{z}\psi^{\prime}\right)$.
By using the identity $\Gamma^{5}\Gamma^{\mu}=\epsilon^{\mu\nu}\Gamma_{\nu}$,
$j^{5\mu}=\left(\epsilon^{\mu\nu}\bar{\psi^{\prime}}\Gamma_{\nu}p_{z}\psi^{\prime}\right)\left(\frac{\tilde{T}_{B}^{3}}{2\pi}\right)$,
where, up to equations of motion, $\bar{\psi^{\prime}}\Gamma_{\nu}p_{\mu}\psi^{\prime}$ is the canonical energy-momentum tensor from the Noether theorem. Hence, we
shall calculate $T_{\mu\nu}$ to obtain $j^{5\mu}$, i.e.,
\begin{equation}
j^{5\mu}=\epsilon^{\mu\nu}T_{\nu3}\left(\frac{\tilde{T}_{B}^{3}}{2\pi}\right),\label{eq:chiral_current-1}
\end{equation}
where $\mu,\thinspace\nu=0,\thinspace3$. Since we are most interested
in the finite-temperature chiral current, we compactify the temporal
direction to a circle of radius $\beta=T^{-1}$. By using the Schwarzian
derivative, one can obtain \citep{francesco2012springer}

\begin{equation}
T_{33}=\frac{c\pi T^{2}}{6},\thinspace\thinspace T_{00}=\frac{c\pi T^{2}}{6},\label{eq:T_33, T_44}
\end{equation}
where $c$ is the central charge. Roughly speaking, $c$ counts the
degrees of freedom. To see this, let us compactify the spatial direction
(to a circle of radius $T^{-1}$) instead and the absolute value of
the energy density is $\left|T \sum_{n}\left(2\pi nT\right)\right|=\frac{\pi T^{2}}{6}$,
which means that each independent mode will contribute a factor $\frac{\pi T^2}{6}$
to the energy density and thus $c$ counts the number of different
modes. By inserting Eq. (\ref{eq:T_33, T_44}) back into Eq. (\ref{eq:chiral_current-1}),
\[
j^{50}=\left(\frac{cT^{2}}{12}\right)\tilde{T}_{B}^{3},
\]
which is exactly the chiral torsional effect we obtained in Eq.
(\ref{eq:chiral_current}).  This can be recast in a covariant form
as
\begin{equation}
j^{5\mu}=-\left(\frac{cT^{2}}{12}\right)\frac{\epsilon^{\mu\nu\rho\sigma}}{\sqrt{|g|}}\eta_{ab}\delta_{\nu}^{a}\partial_{\rho}e_{\sigma}^{b},
\end{equation}
and thus, the chiral anomaly is
\begin{equation}
\frac{1}{\sqrt{|g|}}\partial_{\mu}\sqrt{|g|}j^{5\mu}=-\left(\frac{cT^{2}}{12}\right)\frac{\epsilon^{\mu\nu\rho\sigma}}{\sqrt{|g|}}\partial_{\mu}e_{\nu}^{a}\partial_{\rho}e_{\sigma}^{b}\eta_{ab},\label{eq:thermal_nieh_yan}
\end{equation}
which is similar to the results obtained in the last section. Since the expectation value of the energy-momentum tensor is known to be the free energy density \citep{francesco2012springer}, we have shown explicitly that the prefactor of the Nieh-Yan anomaly and the chiral torsional effect is the free energy density. Interestingly,
we have shown that both the chiral torsional effect and the thermal
Nieh-Yan anomaly are proportional to the central charge. 

Compared to our results in Eq. (\ref{eq:chiral_current}), terms proportional
to cutoff do not appear. This is because the vacuum energy is
from the normal ordering of the creation and annihilation operators,
which is secretly thrown away in conformal field theory due to
the constraints of translational symmetry and rotational symmetry.
However, in realistic materials, both the translational symmetry and
the rotational symmetry can be broken by the ultra-violet physics.
This means that the $\Lambda^{2}$ term might exist in the Nieh-Yan
anomaly, but it is not universal and depends on the concrete systems.

\section{Anomalous thermal Hall effect in Weyl semimetals \label{sec:responses}}

In this section, we shall apply the thermal Nieh-Yan anomaly to Weyl
semimetals. The anomalous thermal Hall effect naturally arises as
the experimental signature of the thermal Nieh-Yan anomaly. The anomalous
thermal Hall conductance is then shown to be proportional to the central
charge.

The anomaly equation in Eq. (\ref{eq:thermal_nieh_yan}) implies that
$\langle j^{5\mu}\rangle=-\frac{cT^{2}}{12}\frac{\epsilon^{\mu\nu\rho\sigma}}{\sqrt{|g|}}e_{\nu}^{a}\partial_{\rho}e_{\sigma}^{b}\eta_{ab}$.
Hence, the effective action in Weyl semimetals can be written as
\begin{eqnarray}
S_{\text{eff}} & = & -\int d^{4}x \sqrt{|g|}b_{\mu}\langle j^{5\mu}\rangle+\dots\nonumber \\
 & = & \frac{cT^{2}}{12}\int d^{4}x \epsilon^{\mu\nu\rho\sigma}b_{\mu}e_{\nu}^{a}\partial_{\rho}e_{\sigma}^{b}\eta_{ab}+\dots,
\end{eqnarray}
where $b^{\mu}$ is the separation between Weyl nodes in the energy-momentum
space and we have kept only terms with linear dependence on $b$.
By performing variation of the vierbeins, the energy-momentum response current is given as 
\begin{equation}
T_{a}^{\mu}=-\frac{cT^{2}}{6}\eta_{ab}\frac{\epsilon^{\mu\nu\rho\sigma}}{\sqrt{|g|}}b_{\nu}\partial_{\rho}e_{\sigma}^{b}.\label{eq:energy_momentum_responses}
\end{equation}
Especially, for $a=0$, $T_{0}^{\mu}$ is the energy current, which
is given as
\begin{equation}
\boldsymbol{j}_{T}=\frac{cT}{6}\boldsymbol{b}\times\left(\boldsymbol{\nabla}T\right).
\end{equation}
 Here $\left(\boldsymbol{j}_{T}\right)^{i}=T_{0}^{i}$ is the thermal current
and we have used $\partial_{0}e_{i}^{0}-\partial_{i}e_{0}^{0}=\left(T^{-1}\boldsymbol{\nabla}T\right)_{i}$
\citep{shitade2014ptep,tatara2015prl}. This is the anomalous thermal Hall effect
in Weyl semimetals and it is shown to be proportional to the central charge, which matches exactly that calculated from the Kubo formula by using the lattice model in the low temperature limit \citep{gorbar2017prb}. Because
the central charge closely relates to the conformal anomaly in
two-dimensional spacetime, the anomalous thermal Hall effect in Weyl
semimetals is thus expected to be protected by topology. In addition, compared to the anomalous quantum Hall effect in Weyl semimetals, i.e.,  $\boldsymbol{j}=\frac{1}{2\pi^2}\boldsymbol{b}\times \boldsymbol{E}$, the ratio of the anomalous thermal Hall conductivity and the anomalous Hall conductivity is $\frac{c\pi^2T}{3}$, which matches the Wiedemann-Franz law exactly.

Equation (\ref{eq:energy_momentum_responses}) also tells
us how the temperature affects the momentum transport in dislocations,
i.e., $\boldsymbol{j}_{p_{m}}=\frac{cT^{2}}{6}b_{0}\nabla\times\boldsymbol{e}^{m},$
where $b_{0}$ is the chiral chemical potential, $\left(\boldsymbol{j}_{p_{m}}\right)^{i}=T_{m}^{i}$
is the $p_{m}$ momentum current and $\left(\boldsymbol{e}^{m}\right)_{i}=-e_{\mu=i}^{a=m}$.
Thus, the total momentum transferred along a dislocation is
\[
J_{p_{m}}=\int_{M}\left(d\boldsymbol{S}\right)\cdot\boldsymbol{j}_{p_{m}}=-\frac{cT^{2}b_{0}}{6}b_{\text{bur}}^{m},
\]
where $M$ is a surface area containing dislocations and $\boldsymbol{b}_{\text{bur}}$
is the Burgers vector, i.e., $\oint dx^{\mu}\wedge dx^{\nu}\partial_{\mu}e_{\nu}^{a}=b_{\text{bur}}^{m}$.

\section{Conclusion}
\label{sec:conclusion}

In summary, we have calculated the torsional Landau levels, from
which the Nieh-Yan anomaly is derived. It was shown that the coefficient
of the Nieh-Yan anomaly is the free energy density of the lowest torsional
Landau levels. By projecting the system onto the lowest torsional
Landau levels consisting of the $\left(1+1\right)$-dimensional Dirac fermions,
we related the Nieh-Yan anomaly to the central charge and thus the
conformal anomaly. The anomalous thermal Hall effect in Weyl semimetals
arises as the direct consequence of the thermal Nieh-Yan anomaly,
which is shown to be proporitonal to the central charge and can be
regarded as  the expermental signature of the thermal Nieh-Yan anomaly.
We have clarified the physical mechanism behind the Nieh-Yan anomaly
and revealed the topological nature of the thermal Nieh-Yan anomaly.

For time-reversal symmetry protected topological insulators, we assume that the negative-mass insulators are the topologically non-trivial ones. The effective action can be obtained by performing a chiral transformation to reverse the sign of the mass. Similarly, the  corresponding torsional effective action for these topological insulators can be derived  from our anomaly equation here, i.e., 
\begin{eqnarray}
S_{\text{TI}}&=&\frac{\pi cT^2}{24}\int d^4 x \eta_{ab}\epsilon^{\mu\nu\rho\sigma}\partial_{\mu}e^{a}_{\nu}\partial_{\rho}e^{b}_{\sigma},
\end{eqnarray}
which suggests the existence of the thermal  counterpart of the magnetoelectric effect in topological insulators.

 In addition, the descent relation between the chiral anomaly and the parity anomaly suggests that there is a corresponding thermal parity anomaly in $(2+1)$ dimensions originating from the thermal Nieh-Yan anomaly, i.e., 
\begin{eqnarray}
S_{\text{TH}}&=&\frac{\pi cT^2}{12}\int \eta_{ab}\epsilon^{\mu\nu\rho}e^{a}_{\mu}\partial_{\nu}e^{b}_{\rho},
\end{eqnarray}
which is proportional to the central charge. This can be used to describe the thermal Hall effect and maybe the topological Hall viscosity. 

\section*{Acknowledgment}

The authors wish to thank Y. Li for insightful discussion.
Z.-M. H and M.~S were not directly supported by any funding agency, but this work would
not be possible without resources provided by the Department
of Physics at the University of Illinois at Urbana-Champaign. B.~H. was supported by ERC Starting Grant No.~678795 TopInSy.

\bibliographystyle{apsrev4-1}

\end{document}